\documentclass[11pt,a4paper]{article}
\usepackage[hyperref]{acl2020}
\usepackage{times}
\usepackage{latexsym}

\usepackage{multirow}
\usepackage{graphicx}

\usepackage{url}

\aclfinalcopy % Uncomment this line for the final submission
\title{An Evaluation of Two Commercial Deep Learning-Based Information Retrieval Systems for COVID-19 Literature}

\author{Sarvesh Soni, Kirk Roberts \\
	School of Biomedical Informatics \\
	University of Texas Health Science Center at Houston \\
	Houston TX, USA \\
	\texttt{\{sarvesh.soni, kirk.roberts\}@uth.tmc.edu} \\
  }

\date{}

\begin{document}
\maketitle

\begin{abstract}

The COVID-19 pandemic has resulted in a tremendous need for access to the latest scientific information, primarily through the use of text mining and search tools.
This has led to both corpora for biomedical articles related to COVID-19 (such as the CORD-19 corpus \cite{wang2020CORD19Covid19Open}) as well as search engines to query such data.
While most research in search engines is performed in the academic field of information retrieval (IR), most academic search engines--though rigorously evaluated--are sparsely utilized, while major commercial web search engines (e.g., Google, Bing) dominate.
This relates to COVID-19 because it can be expected that commercial search engines deployed for the pandemic will gain much higher traction than those produced in academic labs, and thus leads to questions about the empirical performance of these search tools.
This paper seeks to empirically evaluate two such commercial search engines for COVID-19, produced by Google and Amazon, in comparison to the more academic prototypes evaluated in the context of the TREC-COVID track \cite{roberts2020TRECCOVIDRationaleStructure}.
We performed several steps to reduce bias in the available manual judgments in order to ensure a fair comparison of the two systems with those submitted to TREC-COVID.
We find that the top-performing system from TREC-COVID on bpref metric performed the best among the different systems evaluated in this study on all the metrics.
This has implications for developing biomedical retrieval systems for future health crises as well as trust in popular health search engines.

\end{abstract}

\section{Background and Significance}

There has been a surge of scientific studies related to COVID-19 due to the availability of archival sources as well as the expedited review policies of publishing venues. 
A systematic effort to consolidate the flood of such information content, in the form of scientific articles, along with studies from the past that may be relevant to COVID-19 is being carried out as requested by the White House \cite{wang2020CORD19Covid19Open}.
This effort led to the creation of CORD-19, a dataset of scientific articles related to COVID-19 and the other viruses from the coronavirus family.
One of the main aims for building such a dataset is to bridge the gap between machine learning and biomedical expertise to surface insightful information from the abundance of relevant published content.
The TREC-COVID challenge was introduced to target the exploration of the CORD-19 dataset by gathering the information needs of biomedical researchers \cite{roberts2020TRECCOVIDRationaleStructure, voorhees2020TRECCOVIDConstructingPandemic}.
The challenge involved an information retrieval (IR) task to retrieve a set of ranked relevant documents for a given query.
Similar to the task of TREC-COVID, major technology companies Amazon and Google also developed their own systems for exploring the CORD-19 dataset.

Both Amazon and Google have made recent forays into biomedical natural language processing (NLP).
Amazon launched Amazon Comprehend Medical (ACM) for the developers to process unstructured medical data effectively \cite{kass-hout2018IntroducingMedicalLanguage}.
This motivated several researchers to explore the tool's capability in information extraction \cite{bhatia2019ComprehendMedicalNamed, guzman2020AssessmentAmazonComprehend, heider2020ComparativeAnalysisSpeed}.
Interestingly, the same technology is also incorporated to their search engine for the CORD-19 dataset.
It will be useful to assess the overall performance of their search engine that utilizes the company's NLP technology.
Similarly, BERT from Google \cite{devlin2019BERTPretrainingDeep} is enormously popular.
BERT is a powerful language model that is trained on large raw text datasets to learn the nuances of natural language in an efficient manner.
The methodology of training BERT helps it transfer the knowledge from vast raw data sources to other specific domains such as biomedicine.
Several works have explored the efficacy of BERT models in the biomedical domain for tasks such as information extraction \cite{wu2020DeepLearningClinical} and question answering \cite{soni2020EvaluationDatasetSelection}. Many biomedical and scientific variants of the model have also been built, such as BioBERT \cite{lee2019BioBERTPretrainedBiomedical}, Clinical BERT \cite{alsentzer2019PubliclyAvailableClinical}, and SciBERT \cite{beltagy2019SciBERTPretrainedLanguage}.
Google has even incorporated BERT into their web search engine \cite{nayak2019UnderstandingSearchesBetter}.
Since this is the same technology that powers Google's CORD-19 search explorer, it will be interesting to assess the performance of this search tool.

However, despite the popularity of these companies' products, no formal evaluation of these systems is made available by the companies.
Also, neither of these companies participated in the TREC-COVID challenge.
In this paper, we aim to evaluate these two IR systems and compare against the runs submitted to TREC-COVID challenge to gauge the efficacy of what are likely high-utilized search engines.

\section{Methods}
\label{sect:methods}

\begin{table*}[t!]
\caption{Three example topics from Round 1 of the TREC-COVID challenge.} \label{tab:eg-topics}
\centering
\begin{tabular}{|c|r@{\hskip 0.05in} p{0.79\linewidth}|}
\hline
\multirow{4}{*}{\rotatebox{90}{\textbf{Topic 7}}} & \textbf{Query :} & serological tests for coronavirus \\
& \textbf{Question :} & are there serological tests that detect antibodies to coronavirus? \\
& \textbf{Narrative :} & looking for assays that measure immune response to coronavirus that will help determine past infection and subsequent possible immunity. \\
\hline 
\multirow{4}{*}{\rotatebox{90}{\textbf{Topic 10}}} & \textbf{Query :} & coronavirus social distancing impact \\
& \textbf{Question :} & has social distancing had an impact on slowing the spread of COVID-19? \\
& \textbf{Narrative :} & seeking specific information on studies that have measured COVID-19's transmission in one or more social distancing (or non-social distancing) approaches. \\
\hline 
\multirow{4}{*}{\rotatebox{90}{\textbf{Topic 30}}} & \textbf{Query :} & coronavirus remdesivir \\
& \textbf{Question :} & is remdesivir an effective treatment for COVID-19? \\
& \textbf{Narrative :} & seeking specific information on clinical outcomes in COVID-19 patients treated with remdesivir. \\
\hline 
\end{tabular}
\end{table*}

\subsection{Information Retrieval Systems}

We evaluate two publicly available IR systems targeted toward exploring the COVID-19 Open Research Dataset (CORD-19)\footnote{\url{https://www.semanticscholar.org/cord19}} \cite{wang2020CORD19Covid19Open}. These systems are launched by Amazon (CORD-19 Search\footnote{\url{https://cord19.aws}}) and Google (COVID-19 Research Explorer\footnote{\url{https://covid19-research-explorer.appspot.com}}). We hereafter refer to these systems by the names of their corporations, i.e., Amazon and Google. Both the systems take as input a query in the form of natural language and return a list of documents from the CORD-19 dataset ranked by their relevance to the given query.

Amazon’s system uses an enriched version of the CORD-19 dataset constructed by passing it through a language processing service called Amazon Comprehend Medical (ACM) \cite{kass-hout2020AWSLaunchesMachine}.
ACM is a machine learning-based natural language processing (NLP) pipeline to extract clinical concepts such as signs, symptoms, diseases, and treatments from unstructured text \cite{kass-hout2018IntroducingMedicalLanguage}.
The data is further mapped to clinical topics related to COVID-19 such as immunology, clinical trials, and virology using multi-label classification and inference models.
After the enrichment process, the data is indexed using Amazon Kendra that also uses machine learning to provide natural language querying capabilities for extracting relevant documents.

Google’s system is based on a semantic search mechanism powered by BERT \cite{devlin2019BERTPretrainingDeep}, a deep learning-based approach to pre-training and fine-tuning for downstream NLP tasks (document retrieval in this case) \cite{hall2020NLUPoweredToolExplore}.
Semantic search, unlike lexical term-based search that aims at phrasal matching, focuses on understanding the meaning of user queries for searching.
However, deep learning models such as BERT require a substantial amount of annotated data to be tuned for some specific task/domain.
Biomedical articles have very different linguistic features than the general domain, upon which the BERT model is built.
Thus, the model needs to be tuned for the target domain, i.e., biomedical domain, using annotated data.
For this purpose, they use biomedical IR datasets from the BioASQ challenges\footnote{\url{http://bioasq.org}}.
Due to the smaller size of these biomedical datasets, and the large data requirement of the neural models, they use a synthetic query generation technique to augment the existing biomedical IR datasets \cite{ma2020ZeroshotNeuralRetrieval}.
Finally, these expanded datasets are used to fine-tune the neural model.
They further enhance their system by combining term- and neural-based retrieval models by balancing the memorization and generalization dynamics \cite{jiang2020CharacterizingStructuralRegularities}.

\subsection{Evaluation}

We use a topic set collected as part of the TREC-COVID challenge for our evaluations \cite{roberts2020TRECCOVIDRationaleStructure, voorhees2020TRECCOVIDConstructingPandemic}. These topics are a set of information need statements motivated by searches submitted to the National Library of Medicine and suggestions from researchers on Twitter. Each topic consists of three fields with varying levels of granularity in terms of expressing the information need, namely, (a keyword-based) query, (a natural language) question, and (a longer descriptive) narrative. A few example topics from Round 1 of the challenge are presented in Table \ref{tab:eg-topics}. The challenge participants are required to return a ranked list of documents for each topic (also known as runs). The first round of TREC-COVID used a set of 30 topics and exploited the April 10, 2020 release of CORD-19.
Round 1 of the challenge was initiated on April 15, 2020 with the runs from participants due April 23.
Relevance judgments were released May 3.

We use the question and narrative fields from the topics to query the systems developed by Amazon and Google.
These fields are chosen following the recommendations set forward by the organizations, i.e., to use fully formed queries with questions and context.
We use two variations for querying the systems.
In the first variation, we query the systems using only the question.
In the second variation, we also append the narrative to provide more context.

As we accessed these systems in the first week of May 2020, the systems could be using the latest version of CORD-19 at that time (i.e., May 1 release). Thus, we filter the list of returned documents and only include the ones from the April 10 release to ensure a fair comparison with the submissions to the Round 1 of TREC-COVID challenge. We compare the performance of these systems (by Amazon and Google) with the 5 top submissions to the TREC-COVID challenge Round 1 (on the basis of bpref scores). It is valid to compare Amazon and Google systems with the submissions from Round 1 because all these systems are similarly built without using any relevance judgments from TREC-COVID.

Relevance judgments (or assessments) for TREC-COVID are carried out by individuals with biomedical expertise. 
The assessments are performed using a pooling mechanism where only the top-ranked results from different submissions are assessed. 
A document is assigned one of the three possible judgments, namely, \textit{relevant}, \textit{partially relevant}, or \textit{not relevant}. 
We use relevance judgments from Rounds 1 and 2. 
However, even the combined judgments from both the rounds may not ensure that the relevance judgments for top-n documents for both the evaluated systems exist. 
It has recently been shown that pooling effects can negatively impact post-hoc evaluation of systems that did not participate in the pooling \cite{yilmaz2020ReliabilityTestCollections}.
So, to create a level ground for comparison, we perform additional relevance assessments for the documents from evaluated systems that may not have been covered by the combined set of judgments from TREC-COVID. 
In total, 141 documents were assessed by 2 individuals who are also involved in performing the relevance judgments for TREC-COVID.

\begin{table*}[t]
	\caption{Evaluation results after setting a threshold at the number of documents per topic using a minimum number of documents present for each individual topic. The relevance judgments used are a combination of Rounds 1 and 2 of TREC-COVID and our additional relevance assessments. The highest scores for the evaluated and TREC-COVID systems are \underline{underlined}.}
	\label{tab:system-eval-results}
	\centering
	\resizebox{\linewidth}{!}{
	\bgroup
    \def\arraystretch{1.15}
	\begin{tabular}{ccccccccc}
    \hline
     \multicolumn{3}{c}{\textbf{System}} &    \textbf{P@5} & \textbf{P@10} &   \textbf{NDCG@10} &    \textbf{MAP} &   \textbf{NDCG} &   \textbf{bpref} \\
    \hline
    \multirow{2}{*}{Amazon}   & \multicolumn{2}{c}{question}                        & 0.6733 & 0.6333 &     0.539  & 0.0722 &    0.1838 &  0.1049 \\
                                        & \multicolumn{2}{c}{question + narrative}  & \underline{0.72}   & \underline{0.64}   &     \underline{0.5583} & \underline{0.0766} &    \underline{0.1862} &  0.1063 \\
    \multirow{2}{*}{Google}                 & \multicolumn{2}{c}{question}          & 0.5733 & 0.57   &     0.4972 & 0.0693 &    0.1831 &  \underline{0.1069} \\
                                        & \multicolumn{2}{c}{question + narrative}  & 0.6067 & 0.56   &     0.5112 & 0.0687 &    0.1821 &  0.1054 \\
    \hline
    \multirow{5}{*}{\rotatebox{90}{TREC-COVID \hspace{0.1pt} }} & \multicolumn{2}{l}{\hspace{-10pt} 1. sab20.1.meta.docs}      & \underline{0.78}   & \underline{0.7133} &     \underline{0.6109} & \underline{0.0999} &    \underline{0.2266} &  \underline{0.1352} \\
    & \multicolumn{2}{l}{\hspace{-10pt} 2. sab20.1.merged}         & 0.6733 & 0.6433 &     0.5555 & 0.0787 &    0.1971 &  0.1154 \\
    & \multicolumn{2}{l}{\hspace{-10pt} 3. UIowaS\_Run3}           & 0.6467 & 0.6367 &     0.5466 & 0.0952 &    0.2091 &  0.1279 \\
    & \multicolumn{2}{l}{\hspace{-10pt} 4. smith.rm3}              & 0.6467 & 0.6133 &     0.5225 & 0.0914 &    0.2095 &  0.1303 \\
    & \multicolumn{2}{l}{\hspace{-10pt} 5. udel\_fang\_run3}       & 0.6333 & 0.6133 &     0.5398 & 0.0857 &    0.1977 &  0.1187 \\
    \hline
    \end{tabular}
    \egroup}
\end{table*}

The runs submitted to TREC-COVID could contain up to 1000 documents per topic. Due to the restrictions posed by the evaluated systems, we could only fetch up to 100 documents per query. This number further decreases when we remove the documents that are not covered as part of the April 10 release of CORD-19. Thus, to ensure a fair comparison of the evaluated systems with the runs submitted to TREC-COVID, we calculate the minimum number of documents per topic (we call it topic-minimum) across the different variations of querying the evaluated systems (i.e., question or question+narrative). We then use this topic-minimum as a threshold for the maximum number of documents per topic for all evaluated systems.
This ensures that each system returns the same number of documents for a particular topic.

\begin{figure}[t]
    \centering
    \includegraphics[width=1\linewidth]{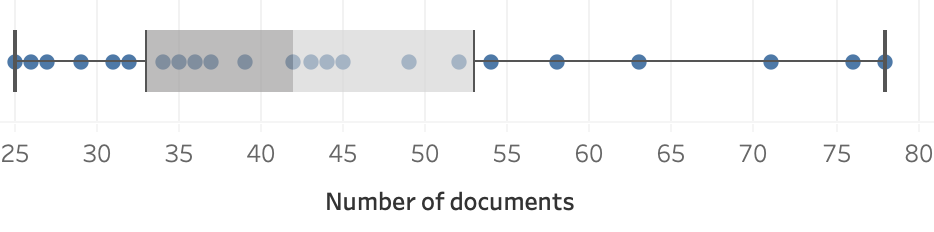}
    \caption{A box plot of the number of documents for each topic as used in our evaluations (after filtering the documents based on the April 10\textsuperscript{th} release of the CORD-19 dataset and setting a threshold at the minimum number of documents for any given topic).}
    \label{fig:num-docs-plot}
\end{figure}

We use the standard measures in our evaluation as employed for TREC-COVID, namely, bpref (binary preference), NDCG@10 (normalized discounted cumulative gain with top 10 documents), and P@5 (precision at 5 documents). Here, bpref only uses judged documents in calculation while the other two measures assume the non-judged documents to be \textit{not relevant}. Additionally, we also calculate MAP (mean average precision), NDCG, and P@10. Note that we can precisely calculate some of the measures that cut the number of documents at up to 10 since we have ensured that both the evaluated systems (for both the query variations) have their top 10 documents manually judged (through TREC-COVID judgments and our additional assessments as part of this study). We use the trec\_eval tool\footnote{\url{https://github.com/usnistgov/trec_eval}} for our evaluations, which is a standard system employed for the TREC challenges.

\section{Results}
\label{sect:results}

The total number of documents used for each topic based on the topic-minimums are shown in the form of a box plot in Figure \ref{fig:num-docs-plot}.
Approximately, an average of 43 documents are evaluated per topic with a median number of documents as 40.5.
This is another reason for using a topic-wise minimum rather than cutting off all the systems to the same level as the lowest return count (that would be 25 documents).
Having a topic-wise cut-off allowed us to evaluate the runs with the maximum possible documents while keeping the evaluation fair.

The evaluation results of our study are presented in Table \ref{tab:system-eval-results}.
Among the commercial systems that we evaluated as part of this study, the question plus narrative variant of the system by Amazon performed consistently better than any other variant in terms of all the included measures other than bpref.
In terms of bpref, the question-only variant of the system from Google performed the best among the evaluated systems.
Note that the best run from the TREC-COVID challenge, after cutting off using topic-minimums, still performed better than the other four submitted runs included in our evaluation.
Interestingly, this best run also performed substantially better than all the variants of both commercial systems evaluated as part of the study on all the calculated metrics.
We discuss more about this system below.

\section{Discussion}
\label{sec:discussion}

We evaluate two commercial IR systems targeted toward extracting relevant documents from the CORD-19 dataset.
For comparison, we also include the 5 best runs from TREC-COVID in our evaluation.
We additionally annotate a total of 141 documents from the runs by the commercial systems to ensure a fair comparison between these runs and the runs from TREC-COVID challenge.
We find that the best system from TREC-COVID in terms of bpref metric outperformed all the commercial system variants on all the evaluated measures including P@5, NDCG@10, and bpref, which are the standard measures used in TREC-COVID.

The commercial systems often employ cutting edge technologies, such as ACM and BERT used by Amazon and Google, while developing their systems.
Also, the availability of technological resources such as CPUs and GPUs may be better in industry settings than in academic settings.
This follows a common concern in academia, namely that the resource requirements for advanced machine learning methods (e.g., GPT-3 \cite{brown2020LanguageModelsAre}) are well beyond the capabilities available to the vast majority of researchers. However, instead these results demonstrate the potential pitfalls of deploying a deep learning-based system without proper tuning. The sabir (sab20.*) system does not use machine learning at all: it is based on the very old SMART system \cite{buckley1985ImplementationSMARTInformation} and does not utilize any biomedical resources.
It is instead carefully deployed based on an analysis of the data fields available in CORD-19.
Subsequent rounds of TREC-COVID have since overtaken sabir (based indeed on machine learning with relevant training data). The lesson, then, for future emerging health events is that deploying ``state-of-the-art'' methods without event-specific data may be dangerous, and in the face of uncertainty simple may still be best.

As evident from Figure \ref{fig:num-docs-plot}, many of the documents retrieved by the commercial systems were not part of the April 10 release of CORD-19.
We queried these systems after another version of the CORD-19 dataset was released.
New sources of papers were constantly being added to the dataset alongside updating the content of existing papers and adding newly published research related to COVID-19.
This may have led to the retrieval of more articles from the new release of the dataset.
However, for a fair comparison between the commercial and the TREC-COVID systems, we pruned the list of documents and performed additional relevance judgments.
We have included the evaluation results that would have resulted without our modifications in the supplemental material.
The performance of these two systems drops precipitously.
Yet, as addressed, this would not have been a ``fair'' comparison and thus the corrective measures described above were necessary to ensure the scientific validity of our comparison.

\section{Conclusion}
\label{sec:conclusion}

We assessed the performance of two commercial IR systems using similar evaluation methods and measures as the TREC-COVID challenge.
To facilitate a fair comparison between these systems and the top 5 runs submitted to the TREC-COVID, we cut all the runs at different thresholds and performed more relevance judgments beyond the assessments provided by TREC-COVID.
We found that the top performing system from TREC-COVID on bpref metric remained the best performing system among the commercial and the TREC-COVID submissions on all the evaluation metrics.
Interestingly, this best performing run comes from a simple system that is purely based on the data elements present in the CORD-19 dataset and does not apply machine learning.
Thus, applying cutting edge technologies without enough target data-specific modifications may not be sufficient for achieving optimal results.

\section*{Acknowledgments}

The authors thank Meghana Gudala and Jordan Godfrey-Stovall for conducting the additional retrieval assessments.
This work was supported in part by the National Science Foundation (NSF) under award OIA-1937136.

\bibliography{acl2020}

\begin{thebibliography}{21}
\expandafter\ifx\csname natexlab\endcsname\relax\def\natexlab#1{#1}\fi

\bibitem[{Alsentzer et~al.(2019)Alsentzer, Murphy, Boag, Weng, Jindi, Naumann,
  and McDermott}]{alsentzer2019PubliclyAvailableClinical}
Emily Alsentzer, John Murphy, William Boag, Wei-Hung Weng, Di~Jindi, Tristan
  Naumann, and Matthew McDermott. 2019.
\newblock \href {https://doi.org/10.18653/v1/W19-1909} {Publicly {{Available
  Clinical BERT Embeddings}}}.
\newblock In \emph{Proceedings of the 2nd {{Clinical Natural Language
  Processing Workshop}}}, pages 72--78.

\bibitem[{Beltagy et~al.(2019)Beltagy, Lo, and
  Cohan}]{beltagy2019SciBERTPretrainedLanguage}
Iz~Beltagy, Kyle Lo, and Arman Cohan. 2019.
\newblock \href {https://doi.org/10.18653/v1/D19-1371} {{{SciBERT}}: {{A
  Pretrained Language Model}} for {{Scientific Text}}}.
\newblock In \emph{Proceedings of the 2019 {{Conference}} on {{Empirical
  Methods}} in {{Natural Language Processing}} and the 9th {{International
  Joint Conference}} on {{Natural Language Processing}}
  ({{EMNLP}}-{{IJCNLP}})}, pages 3615--3620.

\bibitem[{Bhatia et~al.(2019)Bhatia, Celikkaya, Khalilia, and
  Senthivel}]{bhatia2019ComprehendMedicalNamed}
Parminder Bhatia, Busra Celikkaya, Mohammed Khalilia, and Selvan Senthivel.
  2019.
\newblock \href {https://doi.org/10.1109/ICMLA.2019.00297} {Comprehend
  {{Medical}}: {{A Named Entity Recognition}} and {{Relationship Extraction Web
  Service}}}.
\newblock In \emph{2019 18th {{IEEE International Conference On Machine
  Learning And Applications}} ({{ICMLA}})}, pages 1844--1851.

\bibitem[{Brown et~al.(2020)Brown, Mann, Ryder, Subbiah, Kaplan, Dhariwal,
  Neelakantan, Shyam, Sastry, Askell, Agarwal, {Herbert-Voss}, Krueger,
  Henighan, Child, Ramesh, Ziegler, Wu, Winter, Hesse, Chen, Sigler, Litwin,
  Gray, Chess, Clark, Berner, McCandlish, Radford, Sutskever, and
  Amodei}]{brown2020LanguageModelsAre}
Tom~B. Brown, Benjamin Mann, Nick Ryder, Melanie Subbiah, Jared Kaplan,
  Prafulla Dhariwal, Arvind Neelakantan, Pranav Shyam, Girish Sastry, Amanda
  Askell, Sandhini Agarwal, Ariel {Herbert-Voss}, Gretchen Krueger, Tom
  Henighan, Rewon Child, Aditya Ramesh, Daniel~M. Ziegler, Jeffrey Wu, Clemens
  Winter, Christopher Hesse, Mark Chen, Eric Sigler, Mateusz Litwin, Scott
  Gray, Benjamin Chess, Jack Clark, Christopher Berner, Sam McCandlish, Alec
  Radford, Ilya Sutskever, and Dario Amodei. 2020.
\newblock \href {http://arxiv.org/abs/2005.14165} {Language {{Models}} are
  {{Few}}-{{Shot Learners}}}.
\newblock \emph{arXiv:2005.14165 [cs]}.

\bibitem[{Buckley(1985)}]{buckley1985ImplementationSMARTInformation}
Chris Buckley. 1985.
\newblock \href
  {https://ecommons.cornell.edu/bitstream/handle/1813/6526/85-686.pdf}
  {Implementation of the {{SMART}} information retrieval system}.
\newblock Technical {{Report}} 85-686, {Cornell University}.

\bibitem[{Devlin et~al.(2019)Devlin, Chang, Lee, and
  Toutanova}]{devlin2019BERTPretrainingDeep}
Jacob Devlin, Ming-Wei Chang, Kenton Lee, and Kristina Toutanova. 2019.
\newblock \href {https://doi.org/10.18653/v1/N19-1423} {{{BERT}}:
  {{Pre}}-training of {{Deep Bidirectional Transformers}} for {{Language
  Understanding}}}.
\newblock In \emph{Proceedings of the {{North American Chapter}} of the
  {{Association}} for {{Computational Linguistics}}: {{Human Language
  Technologies}}}, pages 4171--4186.

\bibitem[{Guzman et~al.(2020)Guzman, Metzger, Aphinyanaphongs, and
  Grover}]{guzman2020AssessmentAmazonComprehend}
Benedict Guzman, Isabel Metzger, Yindalon Aphinyanaphongs, and Himanshu Grover.
  2020.
\newblock \href {https://arxiv.org/abs/2002.00481v1} {Assessment of {{Amazon
  Comprehend Medical}}: {{Medication Information Extraction}}}.

\bibitem[{Hall(2020)}]{hall2020NLUPoweredToolExplore}
Keith Hall. 2020.
\newblock \href
  {http://ai.googleblog.com/2020/05/an-nlu-powered-tool-to-explore-covid-19.html}
  {An {{NLU}}-{{Powered Tool}} to {{Explore COVID}}-19 {{Scientific
  Literature}}}.

\bibitem[{Heider et~al.(2020)Heider, Obeid, and
  Meystre}]{heider2020ComparativeAnalysisSpeed}
Paul~M. Heider, Jihad~S. Obeid, and St{\'e}phane~M. Meystre. 2020.
\newblock \href {https://www.ncbi.nlm.nih.gov/pmc/articles/PMC7233098/} {A
  {{Comparative Analysis}} of {{Speed}} and {{Accuracy}} for {{Three
  Off}}-the-{{Shelf De}}-{{Identification Tools}}}.
\newblock \emph{AMIA Summits on Translational Science Proceedings},
  2020:241--250.

\bibitem[{Jiang et~al.(2020)Jiang, Zhang, Talwar, and
  Mozer}]{jiang2020CharacterizingStructuralRegularities}
Ziheng Jiang, Chiyuan Zhang, Kunal Talwar, and Michael~C. Mozer. 2020.
\newblock \href {http://arxiv.org/abs/2002.03206} {Characterizing {{Structural
  Regularities}} of {{Labeled Data}} in {{Overparameterized Models}}}.
\newblock \emph{arXiv:2002.03206 [cs, stat]}.

\bibitem[{{Kass-Hout} and Snively(2020)}]{kass-hout2020AWSLaunchesMachine}
Taha~A. {Kass-Hout} and Ben Snively. 2020.
\newblock \href
  {https://aws.amazon.com/blogs/publicsector/aws-launches-machine-learning-enabled-search-capabilities-covid-19-dataset/}
  {{{AWS}} launches machine learning enabled search capabilities for
  {{COVID}}-19 dataset}.

\bibitem[{{Kass-Hout} and Wood(2018)}]{kass-hout2018IntroducingMedicalLanguage}
Taha~A. {Kass-Hout} and Matt Wood. 2018.
\newblock \href
  {https://aws.amazon.com/blogs/machine-learning/introducing-medical-language-processing-with-amazon-comprehend-medical/}
  {Introducing medical language processing with {{Amazon Comprehend Medical}}}.

\bibitem[{Lee et~al.(2019)Lee, Yoon, Kim, Kim, Kim, So, and
  Kang}]{lee2019BioBERTPretrainedBiomedical}
Jinhyuk Lee, Wonjin Yoon, Sungdong Kim, Donghyeon Kim, Sunkyu Kim, Chan~Ho So,
  and Jaewoo Kang. 2019.
\newblock \href {https://doi.org/10.1093/bioinformatics/btz682} {{{BioBERT}}: A
  pre-trained biomedical language representation model for biomedical text
  mining}.
\newblock \emph{Bioinformatics}, pages 1--7.

\bibitem[{Ma et~al.(2020)Ma, Korotkov, Yang, Hall, and
  McDonald}]{ma2020ZeroshotNeuralRetrieval}
Ji~Ma, Ivan Korotkov, Yinfei Yang, Keith Hall, and Ryan McDonald. 2020.
\newblock \href {http://arxiv.org/abs/2004.14503} {Zero-shot {{Neural
  Retrieval}} via {{Domain}}-targeted {{Synthetic Query Generation}}}.
\newblock \emph{arXiv:2004.14503 [cs]}.

\bibitem[{Nayak(2019)}]{nayak2019UnderstandingSearchesBetter}
Pandu Nayak. 2019.
\newblock \href
  {https://blog.google/products/search/search-language-understanding-bert/}
  {Understanding searches better than ever before}.

\bibitem[{Roberts et~al.(2020)Roberts, Alam, Bedrick, {Demner-Fushman}, Lo,
  Soboroff, Voorhees, Wang, and Hersh}]{roberts2020TRECCOVIDRationaleStructure}
Kirk Roberts, Tasmeer Alam, Steven Bedrick, Dina {Demner-Fushman}, Kyle Lo, Ian
  Soboroff, Ellen Voorhees, Lucy~Lu Wang, and William~R. Hersh. 2020.
\newblock \href {https://doi.org/10.1093/jamia/ocaa091} {{{TREC}}-{{COVID}}:
  {{Rationale}} and {{Structure}} of an {{Information Retrieval Shared Task}}
  for {{COVID}}-19}.
\newblock \emph{Journal of the American Medical Informatics Association}.

\bibitem[{Soni and Roberts(2020)}]{soni2020EvaluationDatasetSelection}
Sarvesh Soni and Kirk Roberts. 2020.
\newblock \href {https://www.aclweb.org/anthology/2020.lrec-1.679} {Evaluation
  of {{Dataset Selection}} for {{Pre}}-{{Training}} and {{Fine}}-{{Tuning
  Transformer Language Models}} for {{Clinical Question Answering}}}.
\newblock In \emph{Proceedings of the {{LREC}}}, pages 5534--5540.

\bibitem[{Voorhees et~al.(2020)Voorhees, Alam, Bedrick, {Demner-Fushman},
  Hersh, Lo, Roberts, Soboroff, and
  Wang}]{voorhees2020TRECCOVIDConstructingPandemic}
Ellen Voorhees, Tasmeer Alam, Steven Bedrick, Dina {Demner-Fushman}, William~R.
  Hersh, Kyle Lo, Kirk Roberts, Ian Soboroff, and Lucy~Lu Wang. 2020.
\newblock \href {http://sigir.org/wp-content/uploads/2020/06/p03.pdf}
  {{{TREC}}-{{COVID}}: {{Constructing}} a {{Pandemic Information Retrieval Test
  Collection}}}.
\newblock \emph{ACM SIGIR Forum}, 54:1--12.

\bibitem[{Wang et~al.(2020)Wang, Lo, Chandrasekhar, Reas, Yang, Eide, Funk,
  Kinney, Liu, Merrill, Mooney, Murdick, Rishi, Sheehan, Shen, Stilson, Wade,
  Wang, Wilhelm, Xie, Raymond, Weld, Etzioni, and
  Kohlmeier}]{wang2020CORD19Covid19Open}
Lucy~Lu Wang, Kyle Lo, Yoganand Chandrasekhar, Russell Reas, Jiangjiang Yang,
  Darrin Eide, Kathryn Funk, Rodney Kinney, Ziyang Liu, William Merrill, Paul
  Mooney, Dewey Murdick, Devvret Rishi, Jerry Sheehan, Zhihong Shen, Brandon
  Stilson, Alex~D. Wade, Kuansan Wang, Chris Wilhelm, Boya Xie, Douglas
  Raymond, Daniel~S. Weld, Oren Etzioni, and Sebastian Kohlmeier. 2020.
\newblock \href {https://www.ncbi.nlm.nih.gov/pmc/articles/PMC7251955/}
  {{{CORD}}-19: {{The Covid}}-19 {{Open Research Dataset}}}.
\newblock \emph{arXiv:2004.10706v2}.

\bibitem[{Wu et~al.(2020)Wu, Roberts, Datta, Du, Ji, Si, Soni, Wang, Wei,
  Xiang, Zhao, and Xu}]{wu2020DeepLearningClinical}
Stephen Wu, Kirk Roberts, Surabhi Datta, Jingcheng Du, Zongcheng Ji, Yuqi Si,
  Sarvesh Soni, Qiong Wang, Qiang Wei, Yang Xiang, Bo~Zhao, and Hua Xu. 2020.
\newblock \href {https://doi.org/10.1093/jamia/ocz200} {Deep learning in
  clinical natural language processing: A methodical review}.
\newblock \emph{Journal of the American Medical Informatics Association},
  27:457--470.

\bibitem[{Yilmaz et~al.(2020)Yilmaz, Craswell, Mitra, and
  Campos}]{yilmaz2020ReliabilityTestCollections}
Emine Yilmaz, Nick Craswell, Bhaskar Mitra, and Daniel Campos. 2020.
\newblock \href {https://doi.org/10.1145/3397271.3401317} {On the
  {{Reliability}} of {{Test Collections}} for {{Evaluating Systems}} of
  {{Different Types}}}.
\newblock In \emph{Proceedings of the 43rd {{International ACM SIGIR
  Conference}} on {{Research}} and {{Development}} in {{Information
  Retrieval}}}, pages 2101--2104.

\end{thebibliography}
\bibliographystyle{acl_natbib}

\appendix

\section{Supplementary Material}
\label{sec:supp}

The results without taking into account our additional annotations, i.e., only using the relevance judgments from TREC-COVID rounds 1 and 2, are presented in Table \ref{tab:system-eval-results-without-extra-ann}.
Similarly, the results without setting an explicit threshold on the number of returned documents by the systems are shown in Table \ref{tab:system-eval-results-without-pruning}.
The results without any of the two modifications made by us are provided in Table \ref{tab:system-eval-results-without-extra-ann-and-pruning}.

\begin{table*}[t]
	\caption{Evaluation results after setting a threshold at the number of documents per topic using a minimum number of documents present for each individual topic. The relevance judgments used are a combination of Rounds 1 and 2 of TREC-COVID (WITHOUT our additional relevance assessments). The highest scores for the evaluated and TREC-COVID systems are \underline{underlined}.
	}
	\label{tab:system-eval-results-without-extra-ann}
	\centering
	\resizebox{\linewidth}{!}{
	\bgroup
    \def\arraystretch{1.15}
	\begin{tabular}{ccccccccc}
    \hline
     \multicolumn{3}{c}{\textbf{System}} &    \textbf{P@5} & \textbf{P@10} &   \textbf{NDCG@10} &    \textbf{MAP} &   \textbf{NDCG} &   \textbf{bpref} \\
    \hline
    \multirow{2}{*}{Amazon}   & \multicolumn{2}{c}{question}                        & 0.6467 & \underline{0.5933}  &        0.5095      & 0.069   & 0.1794 &     0.1035 \\
                                        & \multicolumn{2}{c}{question + narrative}  & \underline{0.6933} & \underline{0.5933}  &        \underline{0.5307}      & \underline{0.0722}  & \underline{0.1804} &    0.1031 \\
    \multirow{2}{*}{Google}                 & \multicolumn{2}{c}{question}          & 0.5667 & 0.5133  &        0.4688      & 0.0655  & 0.1785 &     \underline{0.1048} \\
                                        & \multicolumn{2}{c}{question + narrative}  & 0.56   & 0.5133  &        0.4795      & 0.0656  & 0.1763 &    0.1031 \\
    \hline
    \multirow{5}{*}{\rotatebox{90}{TREC-COVID \hspace{0.1pt} }} & \multicolumn{2}{l}{\hspace{-10pt} 1. sab20.1.meta.docs}      & \underline{0.78}   & \underline{0.7133}  &        \underline{0.6109}   & \underline{0.1007}  & \underline{0.2278} &    \underline{0.1361} \\
    & \multicolumn{2}{l}{\hspace{-10pt} 2. sab20.1.merged}         & 0.6667 & 0.64    &        0.5539   & 0.0789  & 0.1968 &    0.1155 \\
    & \multicolumn{2}{l}{\hspace{-10pt} 3. UIowaS\_Run3}           & 0.6467 & 0.6367  &        0.5466       & 0.096   & 0.2099 &   0.1287 \\
    & \multicolumn{2}{l}{\hspace{-10pt} 4. smith.rm3}              & 0.6467 & 0.6133  &        0.5225   & 0.0922  & 0.2107 &    0.1315 \\
    & \multicolumn{2}{l}{\hspace{-10pt} 5. udel\_fang\_run3}       & 0.6333 & 0.6133  &        0.5398       & 0.0866  & 0.1989 &  0.1196 \\
    \hline
    \end{tabular}
    \egroup}
\end{table*}

\begin{table*}[t]
    \vspace{0.11in}
	\caption{Evaluation results WITHOUT setting a threshold at the number of documents per topic using a minimum number of documents present for each individual topic. The relevance judgments used are a combination of Rounds 1 and 2 of TREC-COVID and our additional relevance assessments. The highest scores for the evaluated and TREC-COVID systems are \underline{underlined}.
	}
	\label{tab:system-eval-results-without-pruning}
	\centering
	\resizebox{\linewidth}{!}{
	\bgroup
    \def\arraystretch{1.15}
	\begin{tabular}{ccccccccc}
    \hline
     \multicolumn{3}{c}{\textbf{System}} &    \textbf{P@5} & \textbf{P@10} &   \textbf{NDCG@10} &    \textbf{MAP} &   \textbf{NDCG} &   \textbf{bpref} \\
    \hline
    \multirow{2}{*}{Amazon}   & \multicolumn{2}{c}{question}                        & 0.6733 & 0.6333 &        0.539  & 0.0765 & 0.1931 &  0.1134 \\
                                        & \multicolumn{2}{c}{question + narrative}  & \underline{0.72}   & \underline{0.64}   &        \underline{0.5583} & \underline{0.0788} & 0.1903 &  0.1105 \\
    \multirow{2}{*}{Google}                 & \multicolumn{2}{c}{question}          & 0.5733 & 0.57   &        0.4972 & 0.0775 & \underline{0.2001} &  \underline{0.1227} \\
                                        & \multicolumn{2}{c}{question + narrative}  & 0.6067 & 0.56   &        0.5112 & 0.0763 & 0.1979 &  0.121  \\
    \hline
    \multirow{5}{*}{\rotatebox{90}{TREC-COVID \hspace{0.1pt} }} & \multicolumn{2}{l}{\hspace{-10pt} 1. sab20.1.meta.docs}   & \underline{0.78}   & \underline{0.7133} &        \underline{0.6109} & \underline{0.2037} & \underline{0.4702} &  0.3404 \\
    & \multicolumn{2}{l}{\hspace{-10pt} 2. sab20.1.merged}         															& 0.6733 & 0.6433 &        0.5555 & 0.1598 & 0.4415 &  \underline{0.3433} \\
    & \multicolumn{2}{l}{\hspace{-10pt} 3. UIowaS\_Run3}           															& 0.6467 & 0.6367 &        0.5466 & 0.174  & 0.4145 &  0.3229 \\
    & \multicolumn{2}{l}{\hspace{-10pt} 4. smith.rm3}              															& 0.6467 & 0.6133 &        0.5225 & 0.1947 & 0.4461 &  0.3406 \\
    & \multicolumn{2}{l}{\hspace{-10pt} 5. udel\_fang\_run3}       															& 0.6333 & 0.6133 &        0.5398 & 0.1911 & 0.4495 &  0.3246 \\
    \hline
    \end{tabular}
    \egroup}
\end{table*}

\begin{table*}[t]
    \vspace{0.11in}
	\caption{Evaluation results WITHOUT setting a threshold at the number of documents per topic using a minimum number of documents present for each individual topic. The relevance judgments used are a combination of Rounds 1 and 2 of TREC-COVID (WITHOUT our additional relevance assessments). The highest scores for the evaluated and TREC-COVID systems are \underline{underlined}.
	}
	\label{tab:system-eval-results-without-extra-ann-and-pruning}
	\centering
	\resizebox{\linewidth}{!}{
	\bgroup
    \def\arraystretch{1.15}
	\begin{tabular}{ccccccccc}
    \hline
     \multicolumn{3}{c}{\textbf{System}} &    \textbf{P@5} & \textbf{P@10} &   \textbf{NDCG@10} &    \textbf{MAP} &   \textbf{NDCG} &   \textbf{bpref} \\
    \hline
    \multirow{2}{*}{Amazon}   & \multicolumn{2}{c}{question}                        & 0.6467 & \underline{0.5933} &        0.5095 & 0.0732 & 0.1888 &  0.1121 \\
                                        & \multicolumn{2}{c}{question + narrative}  & \underline{0.6933} & \underline{0.5933} &        \underline{0.5307} & \underline{0.0744} & 0.1846 &  0.1074 \\
    \multirow{2}{*}{Google}                 & \multicolumn{2}{c}{question}          & 0.5667 & 0.5133 &        0.4688 & 0.0734 & \underline{0.1954} &  \underline{0.1208} \\
                                        & \multicolumn{2}{c}{question + narrative}  & 0.56   & 0.5133 &        0.4795 & 0.0728 & 0.1919 &  0.1188 \\
    \hline
    \multirow{5}{*}{\rotatebox{90}{TREC-COVID \hspace{0.1pt} }} & \multicolumn{2}{l}{\hspace{-10pt} 1. sab20.1.meta.docs}   & \underline{0.78}   & \underline{0.7133} &        \underline{0.6109} & \underline{0.2038} & \underline{0.4693} &  0.3406 \\
    & \multicolumn{2}{l}{\hspace{-10pt} 2. sab20.1.merged}         															& 0.6667 & 0.64   &        0.5539 & 0.1589 & 0.4393 &  \underline{0.3426} \\
    & \multicolumn{2}{l}{\hspace{-10pt} 3. UIowaS\_Run3}           															& 0.6467 & 0.6367 &        0.5466 & 0.1742 & 0.4139 &  0.3225 \\
    & \multicolumn{2}{l}{\hspace{-10pt} 4. smith.rm3}              															& 0.6467 & 0.6133 &        0.5225 & 0.1956 & 0.4469 &  0.3413 \\
    & \multicolumn{2}{l}{\hspace{-10pt} 5. udel\_fang\_run3}       															& 0.6333 & 0.6133 &        0.5398 & 0.1914 & 0.4497 &  0.3248 \\
    \hline
    \end{tabular}
    \egroup}
\end{table*}

\end{document}